# ENHANCING PERSONALIZED RECIPE RECOMMENDATION THROUGH MULTI-CLASS CLASSIFICATION

Harish Neelam[1] and Koushik Sai Veerella[2]

[1]Department of Epidemiology and Biostatistics, Michigan State University, East Lansing, Michigan, USA
[2]Department of Electrical and Computer Engineering, Michigan State University, East Lansing, Michigan, USA

## ABSTRACT

*This paper intends to address the challenge of personalized recipe recommendation in the realm of diverse culinary preferences. The problem domain involves recipe recommendations, utilizing techniques such as association analysis and classification. Association analysis explores the relationships and connections between different ingredients to enhance the user experience. Meanwhile, the classification aspect involves categorizing recipes based on user-defined ingredients and preferences. A unique aspect of the paper is the consideration of recipes and ingredients belonging to multiple classes, recognizing the complexity of culinary combinations. This necessitates a sophisticated approach to classification and recommendation, ensuring the system accommodates the nature of recipe categorization. The paper seeks not only to recommend recipes but also to explore the process involved in achieving accurate and personalized recommendations.*

## 1. INTRODUCTION

In the realm of recipe recommendation systems, the focus on ingredient-based recommendations remains limited. This project addresses this gap by employing advanced data mining techniques, specifically association rules, classification-based filtering, and network analysis, to provide a comprehensive and user-specific recipe recommendation system using 10K recipes [1]. Recipe recommendation systems traditionally rely on factors such as cuisine, dietary considerations, or meal courses [2]. However, there is a noticeable scarcity of systems solely based on ingredients. We are driven by the ambition to develop a user-focused recommendation system that suggests recipes based on provided ingredients, catering to individual preferences and dietary needs. The primary objective of this research is to create an ingredient-based recommendation system, employing advanced data mining techniques, and enhance algorithmic robustness and precision through multi-class classification.

The data utilized for this project was gathered through web scraping, resulting in three distinct datasets: Cultural Cuisines, Dietary Considerations, and Meal Course. The inherent challenge lay in the extensive pre-processing required to transform unrefined data into an organized format for subsequent analytical processes. After acquiring data through scraping, we pre-processed ingredients using similarity scores like cosine similarity, inspired by Cohen et al. [3]. To extract meaningful patterns from the data, we applied both Apriori and FP Growth algorithms for association rule mining [4] for robust recommendations based on Sandvig et al. [5] and used advanced classification models informed by Kowsari et al. [6] to categorize ingredients.

For the classification-based filtering aspect, we explored various models, including KNN, Random Forest, SGD, XGBoost, and Multinomial naive bayes. Notably, Stochastic gradient descent classifier (SGD, a linear classifier optimized by the stochastic gradient descent algorithm) proven to be the highest-performing model [7], accomplishing an accuracy of 80.13% in the cultural cuisine dataset, 50.67% in Dietary considerations, and 70.46% in the Meal course dataset.

The primary contributions of this paper include: Implementation and exploration of Apriori and FP Growth algorithms for association rule mining, Evaluation of multiple classification models, with SGD identified as the most effective for predicting classes in cultural cuisines, dietary considerations, and meal courses, Introduction of a user-centric recipe recommendation system based on ingredient analysis, Integration of network analysis to visualize ingredient relationships within recommended recipes [8]. The remainder of this paper unfolds as follows: Section 2 elucidates the related literature regarding this paper. Section 3 elaborates on the methodologies which includes pre-processing steps undertaken to refine the raw data, the strategies employed for association rule-based recommendations, highlighting the significance of Apriori and FP Growth algorithms, Classification-based filtering methodology, emphasizing the performance of the SGD model, the network analysis of ingredients, offering insights into the intricate relationships within recommended recipes. Section 4 illustrates the evaluation results and finally, a conclusion regarding our paper.

## 2. RELATED WORK

After acquiring data through scraping, our focus shifted to pre-processing ingredients. Drawing inspiration from the research of Cohen, Ravikumar, and Fienberg [3] in "A Comparison of String Distance Metrics for Name-Matching Tasks," we explored similarity scores such as Jaro-Winkler and cosine similarity. While the former was recommended, we found that cosine similarity performed reasonably well for our text pre-processing needs.

We delved into the "Robustness of collaborative recommendation through association rule mining", as discussed by Sandvig, Mobasher, and Burke [5]. Their work emphasized the effectiveness of association rule-based recommenders in achieving algorithmic robustness. Consequently, we adopted association rule mining as a fundamental concept in recommending recipes. To further enhance the precision of our recommended recipes, we sought to classify ingredients and determine their respective categories. Building on the insights from Kowsari, Meimandi, Heidarysafa, Mendu, Barnes, and Brown [6] in "Text Classification Algorithms," we implemented sophisticated classification algorithms tailored to our ingredient data.

In their 2010 paper, "Intelligent food planning: personalized recipe recommendation," Freyne, Jill, and Shlomo Berkovsky [9] propose a recipe recommender system to combat the obesity epidemic. Their approach involves focusing on user engagement by exploring data capture, food-recipe relationships, and the adaptability of recommender algorithms for personalized, healthy

International Journal of Computer Science, Engineering and Information Technology (IJCSEIT), Vol.14, No.5, August 2024

recipe suggestions. Our system aims to address the issue by filtering out ingredients known to contribute to obesity, ensuring that recommended recipes align with users' health goals.

The goal of providing users with personalized ingredient networks was inspired by the work of Teng, Lin, and Adamic [10] in "Recipe recommendation using ingredient networks" and Ahnert [11] in "Network analysis and data mining in food science." Our approach distinguishes itself in the limited landscape of prior studies, leveraging these guiding principles while incorporating innovative strategies to develop a prototype for recipe recommendations.

## 3. METHODOLOGY

In the prototype development of our recipe recommendation project, we employed a diverse set of techniques at different stages, encompassing data collection, pre-processing, and modelling.

### 3.1. Data Collection

To construct our dataset, we undertook web scraping to compile recipe information from the RecipeLand website [1]. We aimed to accumulate a wide array of recipes spanning various categories, including those to specific dietary considerations, different cuisines, and distinct dish types. For the web scraping process, we utilized the requests library for handling HTTP requests and the BeautifulSoup library for HTML content parsing [12]. This approach allowed us to systematically extract key information from each recipe, including the recipe title, list of ingredients, and the corresponding category. To address concerns about the data collection methodology, we ensured our web scraping process was thorough and systematic, minimizing potential biases as outlined in studies on complex network sampling issues. We employed robust techniques to improve the reliability of our datasets and diminish errors associated with incomplete or biased sampling [13]. The following figures (Figures 1 and 2) illustrate the average number of Ingredients used in a recipe type and the distribution of recipes by type in the data.

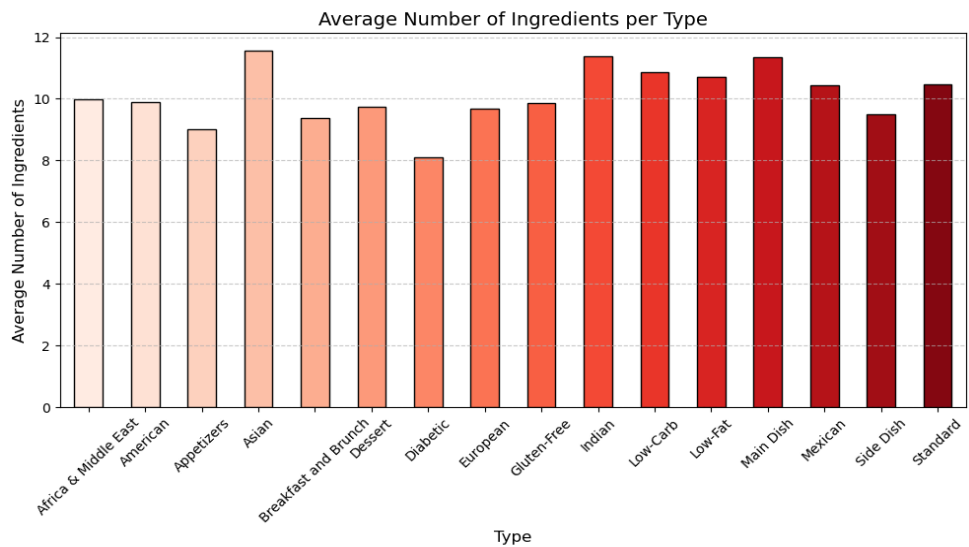

Figure 1. Bar plot depicting Avg. ingredients per Dish Type.

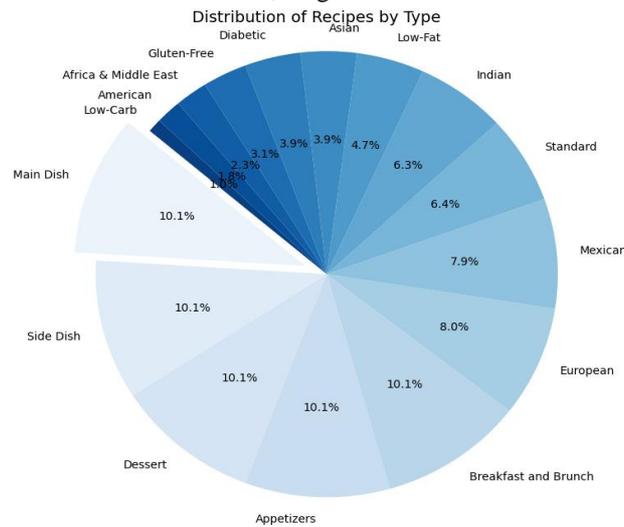

Figure 2. Pie chart of Data Distribution of Recipe types.

## 3.2. Data Pre-Processing

After extracting data from the website, we conducted essential pre-processing steps to refine the data for subsequent stages [14]. Our process began by eliminating stop-words using the NLTK library [15]. Following that, we addressed extra white spaces and non-alphabetic characters in the dataset. A crucial step involved removing quantity and measurement information associated with the recipes, such as terms like "cup" and "teaspoon."

To further enhance data quality, we filtered out empty strings and eliminated parenthetical content. Employing Comma-Based Ingredient Segmentation proved effective in removing irrelevant data from the dataset. Finally, we streamlined the dataset by eliminating duplicate entries, ensuring that it exclusively contained unique recipes. This comprehensive pre-processing pipeline resulted in a more refined and usable dataset for subsequent stages of the project.

## 3.3. Similarity Score

To further refine our dataset, we applied similarity metrics to assess the likeness between ingredients. If the similarity exceeded a specified threshold, we merged the two ingredients into a single entry. For instance, if our dataset contained both "Chicken Breast" and "Chicken Thighs," we consolidated them into a single category—just "Chicken." This simplification facilitates user input, allowing for a more straightforward experience.

To achieve this, we implemented various similarity scores, including Jaccard, Cosine, and Jaro-Winkler Similarities. While Jaccard didn't yield optimal results for our case [3], both Cosine and Jaro-Winkler Similarities proved satisfactory. Although the disparity in performance between the two was not substantial, they effectively captured the majority of similarities in the ingredients.

## 3.4. Association Rule Mining

Next, we used Association Rule Mining on our cleaned-up ingredient list to find the most common ingredient combinations. We tried both the Apriori and FP-growth algorithms [4] from the 'mlxtend' library. Surprisingly, both algorithms gave similar results, but we settled on Apriori.

Given our relatively small dataset, we had to set low values for support (0.02) and threshold (0.2) to get enough item sets. This adjustment was crucial to make sure we captured a good number of useful rules for our recipe recommendations.

### 3.4.1. Strategy for Recommendations

The approach we adopted for recommending recipes based on user-input ingredients is as follows: First, we collect the user's input ingredients into a list. Subsequently, we identify rules for each ingredient in the input list to determine their corresponding consequents. Next, we generate all possible combinations of the antecedents and consequent ingredients. The resulting combinations serve as criteria for recommending recipes that include these specific ingredient combinations.

For instance, if the user provides the ingredients "garlic" and "basil," we seek rules for both ingredients. This process may yield rules such as:

{garlic} → {onions} and {basil} → {tomatoes}
Then we use these rules to find all the combinations of the ingredients in the rules keeping {garlic, basil} as the base.
{garlic, basil}
{garlic, basil, onions}
{garlic, basil, tomatoes}
{garlic, basil, onions, tomatoes}

We recommend only the recipes in which these combinations of ingredients are present in the data set we created.

## 3.5. Classification Based Filtering

To further enhance the results, we implemented classification based on the provided ingredients. We trained distinct classification models on three datasets, aiming to categorize user-specified ingredients into different groups, including Cuisines, Dietary considerations, and Meal courses. The models employed for this task included KNN, Random Forest, SGD, Multinomial NB, and XGBoost.

Following the training phase on the three datasets, we performed hyperparameter tuning to enhance the accuracy of each model [16]. The outcomes highlighted SGD and XGBoost as the most effective models for our approach, exhibiting approximately similar scores. SGD performs well on text classification due to its efficiency with large datasets, ability to handle sparse data, ease of parallelization, adaptability to online learning, and regularization options. Its incremental learning approach and simplicity in implementation make it suitable for text classification tasks. Subsequently, we opted to utilize the SGD model to predict the category to which the ingredients belong [7].

### 3.5.1. Strategy for Classification

After obtaining the results from the classification models, we further refined our classification strategy by employing three separate SGD models. We recorded the probabilities for each class, and with a threshold set at 0.3, we assigned the classes to the ingredients.

Acknowledging that ingredients could potentially belong to multiple classes in the dataset, we recognized that combining all three datasets into a single large dataset would result in a challenging 16-class multiclass classification problem. To address this, we implemented three models for each dataset to improve accuracy and handle the complexity of ingredients belonging to multiple classes simultaneously.

For instance, consider the following ingredients: {garlic, basil, tomatoes, onions}. This set could be associated with multiple classes, such as {Asian, American}, {Main Dish}, and {Low fat, Gluten-free}. This approach allows us to better capture the best categorization of ingredients.

### 3.6. Network Analysis

To enhance further understanding of how recommended recipe ingredients are interconnected, we utilized the NetworkX and PyVis libraries to create visualizations [17] of ingredient networks. This network design aims to illustrate the complex connections between ingredients and various recipe types. The goal is to provide insights derived from the network analysis [10], allowing to customize the recipe recommendations based on a visual representation of ingredient relationships. In Figure 3, three distinct clusters are evident, offering us the flexibility to select the cluster of our interest or fine-tune recipe recommendations by excluding unwanted ingredients.

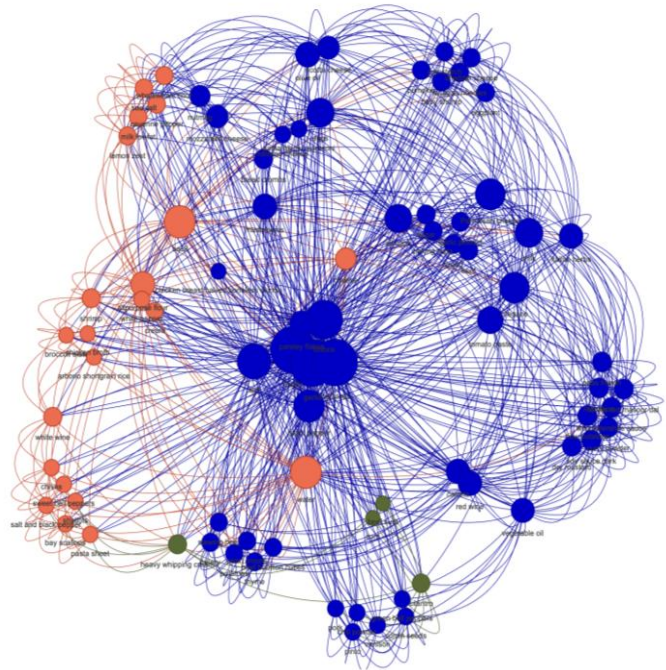

Figure 3. Ingredient Network of Recommended recipes for a given set.

## 4. RESULTS

The dataset consistently features three attributes obtained through web scraping. Initially, our dataset contained extra words, white spaces, non-alphabetic characters, and empty strings. Through pre-processing, we successfully eliminated this unwanted information without losing any instances. While intentional duplicates exist in our dataset, each instance is unique, particularly as some recipes belong to multiple classes.

The evaluation measures that we have used are Confusion matrix, and accuracy to compare the classification models [18] described in the previous sections. The Accuracy scores are shown in the Table 1 as follows:

Table 1: Accuracy results of SGD Classifier

| Dataset | Number of Classes | Accuracy (%) |
|---|---|---|
| **Dietary** | 5 | 50.76 |
| **Cuisines** | 6 | 80.13 |
| **Course** | 5 | 70.46 |

The poor result in the Dietary Dataset is because a single recipe is repeated in multiple classes, which we thought was the ideal case. One of the limitations that we had is less data, with more data, the models could have performed better and would have resulted in more association rules. The confusion matrices of highest achieved accuracy by the SGD classifiers in 3 datasets are shown in the following Figures 4, 5, 6.

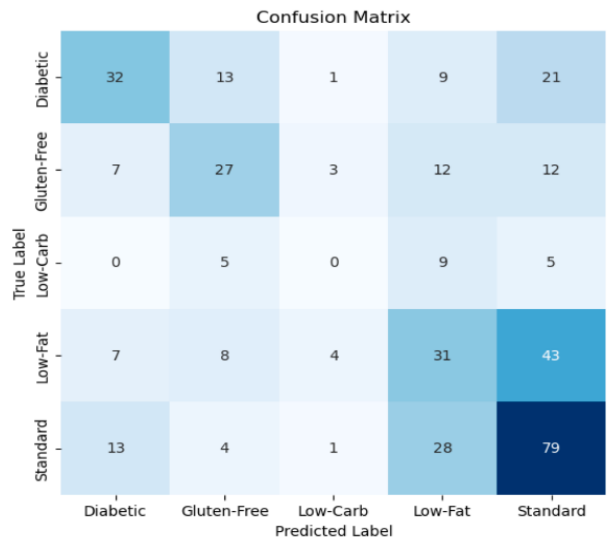

Figure 4. Confusion matrix of Highest Accuracy achieving SGD Classifier in Dietary Dataset.

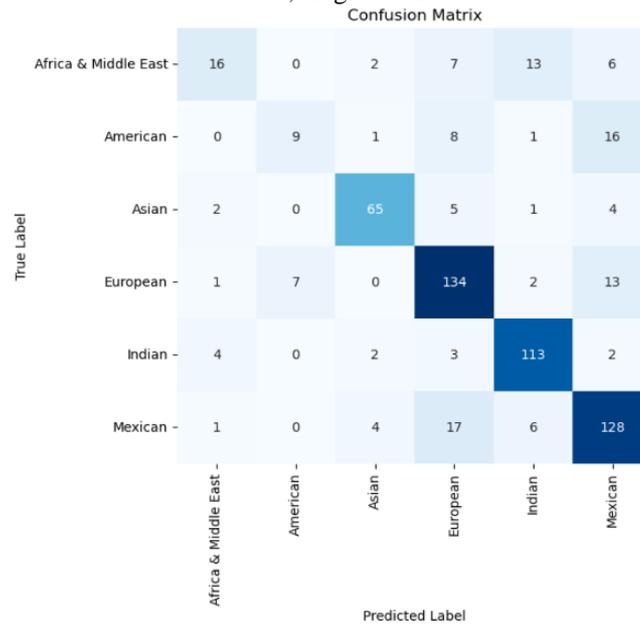

Figure 5. Confusion matrix of Highest Accuracy achieving SGD Classifier in Cuisines Dataset.

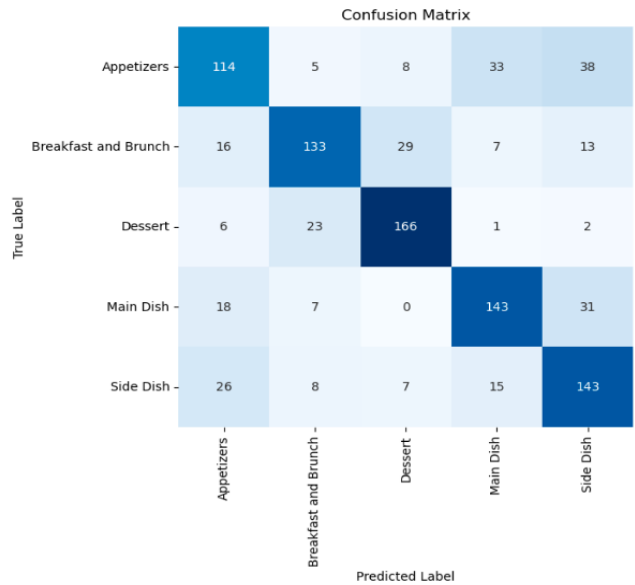

Figure 6. Confusion matrix of Highest Accuracy achieving SGD Classifier in Course Dataset.

## 4.1. Proposed Architecture

Our recommendation system introduces a novel approach focusing on ingredients to provide personalized recipe suggestions. The architecture comprises several interconnected components, including data mining techniques, classification models, and network analysis to deliver tailored recommendations.

**Association Rule Mining:** Apriori Algorithm is employed for association rule mining. This technique analyses the relationships between ingredients within recipes, identifying frequent item sets and association rules that guide the recommendation process. The algorithm recommends recipes containing combinations of ingredients specified by the user and generated rules.

**Classification Model:** Stochastic Gradient Descent (SGD) is used to provide accurate classification of recipes according to user-defined criteria. The algorithm categorizes recipes into classes based on recommended recipes and user preferences, effectively filtering them to match user flavour profiles.

**Network Analysis:** Network analysis is integrated to visualize ingredient relationships within recommended recipes. This component enhances the understanding of underlying connections between ingredients, providing users with insights into recipe composition and flavour profiles. Users can utilize this visualization to select preferred ingredients and manage recommended recipes accordingly.

### 4.2. Discussion

Our study yields valuable insights into the efficacy of our recommendation system. Our system demonstrates promising results in delivering tailored recipe recommendations aligned with user preferences and ingredient specifications. The accuracies for multi-class classification are notably high in our case. With additional data and further training, the models have the potential to perform even better. The strategies mentioned in this paper for extracting rules and filtering significantly enhance the system, resulting in a more refined and suitable list of recipes to recommend.

Compared to traditional recommendation systems, our approach offers a unique focus on ingredients, allowing for more precise and personalized recipe suggestions. Integration of network analysis further enriches the user experience by providing visual insights into ingredient relationships, setting our system apart from others in the field. Future work may involve further improving the system by incorporating additional data sources and expanding the dataset to uncover more intricate ingredient patterns, ultimately enhancing the system's effectiveness and usability.

## 5. CONCLUSIONS

This paper introduces a novel recommendation system that focuses exclusively on ingredients, providing recipe suggestions based on specified ingredients. Application of state-of-the-art data mining techniques, including Apriori and FP Growth algorithms for association rule mining, and the evaluation of multiple classification models, with Stochastic Gradient Descent (SGD) identified as the most effective for classifying recipes. Integration of network analysis to visualize ingredient relationships within recommended recipes, enhancing the understanding of underlying connections between ingredients. In summary, through the combined use of association rule mining and classification techniques, our system delivers tailored recipe recommendations aligned with user preferences and ingredient specifications, facilitating an enriched culinary experience.

Future work may involve incorporating data from additional external sources, such as user reviews, nutritional information, or real-time ingredient availability, to further refine and enrich the recommendation system and collecting a larger and more diverse data to further enrich the association rule mining and classification processes. A more extensive dataset could lead to the discovery of additional patterns within ingredient combinations, ultimately enhancing the accuracy and coverage of the recommendation system.

ACKNOWLEDGEMENTS

Dr. Pang-Ning Tan, Professor of CSE Department, Michigan State University.

**AUTHORS**

**Harish Neelam** has received MS in Data Science from Michigan State University and currently pursuing PhD in Biostatistics at Michigan State University. His Research interests include High-Dimensional data mining, Optimization, Environmental Health, Genetic Epidemiology and Advanced statistical methods.

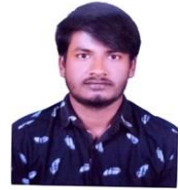

**Koushik Sai Veerella** has received MS in Data Science from Michigan State University and currently pursuing PhD in Electrical and Computer Engineering at Michigan State University. His Research interests include Natural Language Processing, Generative AI and Large Language Models.

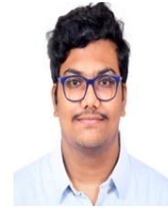